\documentclass[%
reprint,
superscriptaddress,
amsmath,amssymb,
aps,longbibliography,
prb,
floatfix,
]{revtex4-2}

\usepackage{graphicx}%
\usepackage{verbatim}
\usepackage[version=4]{mhchem}
\usepackage{xcolor}
\usepackage{physics}
\usepackage[normalem]{ulem}
\usepackage{pdfpages}
\usepackage{pgffor}

\makeatletter
\AtBeginDocument{\let\LS@rot\@undefined}
\makeatother

\ifdefined\COSMOMATHS
\else
\newcommand{\COSMOMATHS}{}

\newcommand{\mbf}[1]{\ensuremath{\mathbf{#1}}}

\newcommand{\D}[1]{\operatorname{d}{\!#1}\,}

\fi %

\ifdefined\DIRACREP
\else
\newcommand{\DIRACREP}{}

\usepackage{physics}
\usepackage{xparse}
\ifdefined\COSMOMATHS
\else
\newcommand{\COSMOMATHS}{}

\newcommand{\mbf}[1]{\ensuremath{\mathbf{#1}}}

\newcommand{\D}[1]{\operatorname{d}{\!#1}\,}

\fi %

\NewDocumentCommand{\rep}{s d<| d|>}{%
\IfBooleanTF{#1}{
   \IfValueTF{#2}{
       \IfValueTF{#3}{\braket{#2}{#3}}{\bra{#2}}
       }{
       \IfValueTF{#3}{\ket{#3}}{}
       }
   }{
   \IfValueTF{#2}{
       \IfValueTF{#3}{\braket*{#2}{#3}}{\bra*{#2}}
       }{
       \IfValueTF{#3}{\ket*{#3}}{}
       }
   }
}

\NewDocumentCommand{\rbra}{sm}{\IfBooleanTF{#1}{\rep*<#2|}{\rep<#2|}}
\NewDocumentCommand{\rket}{sm}{\IfBooleanTF{#1}{\rep*|#2>}{\rep|#2>}}
\NewDocumentCommand{\rbraket}{smom}{
    \IfBooleanTF{#1}{
        \IfNoValueTF{#3}{\rep*<#2||#4>}{\rep*<#2|#3\rep*|#4>}
    }{
        \IfNoValueTF{#3}{\rep<#2||#4>}{\rep<#2|#3\rep|#4>}
    }
}

\NewDocumentCommand{\cg}{m m m}{\rep<#1; #2||#3>}

\NewDocumentCommand{\field}{o m e{_} e{^} o e{_} e{^}}{
\IfValueTF{#5}{\overline{
  #2\IfValueT{#3}{_#3}\IfValueT{#4}{^{\otimes #4}} %
  \otimes
  #5\IfValueT{#6}{_#6}\IfValueT{#7}{^{\otimes #7}} %
  \IfValueT{#1}{;#1}
}}{
  \IfValueTF{#4}{\overline{
     #2\IfValueT{#3}{_#3}\IfValueT{#4}{^{\otimes #4}}
     \IfValueT{#1}{;#1}
  }}
  {#2\IfValueT{#3}{_#3}}
}
}

\NewDocumentCommand{\frho}{o e{_} e{^}}{
\field[#1]{\rho}_{#2}^{#3}
}

\newcommand{\br}{\mbf{r}}
\newcommand{\bv}{\mbf{v}}
\newcommand{\bx}{\mbf{x}}
\newcommand{\bxhat}{\hat{\mbf{x}}}
\newcommand{\brhat}{\hat{\mbf{r}}}

\newcommand{\e}{a}  %

\NewDocumentCommand{\ex}{e_}{
\IfValueTF{#1}{\e_{#1}\bx_{#1}}{\e\bx}
}  %

\NewDocumentCommand{\lm}{e_}{
\IfValueTF{#1}{l_{#1}m_{#1}}{lm}
}
\NewDocumentCommand{\nlm}{e_}{
\IfValueTF{#1}{n_{#1}\lm_{#1}}{n\lm}
}
\NewDocumentCommand{\enlm}{e_}{
\IfValueTF{#1}{\e_{#1}\nlm_{#1}}{\e\nlm}
}
\NewDocumentCommand{\en}{e_}{
\IfValueTF{#1}{\e_{#1}n_{#1}}{\e n}
}
\NewDocumentCommand{\nl}{e_}{
\IfValueTF{#1}{n_{#1}l_{#1}}{nl}
}
\NewDocumentCommand{\nlk}{e_}{
\IfValueTF{#1}{n_{#1}l_{#1}k_{#1}}{nlk}
}
\NewDocumentCommand{\enlk}{e_}{
\IfValueTF{#1}{\e_{#1}\nlk_{#1}}{\e\nlk}
}
\NewDocumentCommand{\enl}{e_}{
\IfValueTF{#1}{\en_{#1}l_#1}{\en l}
}

\NewDocumentCommand{\nnl}{s}{
\IfBooleanTF{#1}{n_1 n_2 l}{n_1; n_2; l}
}
\NewDocumentCommand{\ennl}{s}{
\IfBooleanTF{#1}{\en_1 \en_2 l}{\en_1; \en_2; l}
}

\NewDocumentCommand{\gslm}{s}{
\IfBooleanTF{#1}{\sigma\lambda\mu}{\sigma;\lambda\mu}
}
\NewDocumentCommand{\glm}{}{\lambda\mu}

\newcommand{\Rhat}{\hat{R}}

\fi %

\ifdefined\COSMOMODELS
\else
\newcommand{\COSMOMODELS}{}
\usepackage{bm,upgreek}

\newcommand{\y}[0]{y}

\newcommand{\by}[0]{\mbf{\y}}

\newcommand{\feat}{\upxi}
\newcommand{\bfeat}[0]{\ensuremath{\bm{\upxi}}}

\fi %

\newcommand{\natoms}{\ensuremath{n_\text{at}}}

\begin{document}

\title{Completeness of Atomic Structure Representations}

\author{Jigyasa Nigam}
\affiliation{Laboratory of Computational Science and Modeling, Institut des Mat\'eriaux, \'Ecole Polytechnique F\'ed\'erale de Lausanne, 1015 Lausanne, Switzerland}

\author{Sergey N. Pozdnyakov}
\affiliation{Laboratory of Computational Science and Modeling, Institut des Mat\'eriaux, \'Ecole Polytechnique F\'ed\'erale de Lausanne, 1015 Lausanne, Switzerland}

\author{Kevin K. Huguenin-Dumittan}
\affiliation{Laboratory of Computational Science and Modeling, Institut des Mat\'eriaux, \'Ecole Polytechnique F\'ed\'erale de Lausanne, 1015 Lausanne, Switzerland}

\author{Michele Ceriotti}
\email{michele.ceriotti@epfl.ch}
\affiliation{Laboratory of Computational Science and Modeling, Institut des Mat\'eriaux, \'Ecole Polytechnique F\'ed\'erale de Lausanne, 1015 Lausanne, Switzerland}

\date{\today}%

\begin{abstract}
In this paper, we address the challenge of obtaining a comprehensive and symmetric representation of point particle groups, such as atoms in a molecule, which is crucial in physics and theoretical chemistry. 
The problem has become even more important with the widespread adoption of machine-learning techniques in science, as it underpins the capacity of models to accurately reproduce  physical relationships while being consistent with fundamental symmetries and conservation laws.
However, some of the descriptors that are commonly used to represent point clouds -- most notably those based on discretized correlations of the neighbor density, that underpin most of the existing ML models of matter at the atomic scale -- are unable to distinguish between special arrangements of particles in three dimensions. 
This makes it impossible to machine learn their properties.
Atom-density correlations are provably complete in the limit in which they simultaneously describe the mutual relationship between all atoms, which is impractical.
We present a novel approach to construct descriptors of \emph{finite} correlations based on the relative arrangement of particle triplets, which can be employed to create symmetry-adapted models with universal approximation capabilities, which have the resolution of the neighbor discretization as the sole convergence parameter. 
Our strategy is demonstrated on a class of atomic arrangements that are specifically built to defy a broad class of conventional symmetric descriptors, showcasing its potential for addressing their limitations.

\end{abstract}

\maketitle

The construction of data-driven models of physics starts by defining a suitable mathematical representation of the problem, which is then used as the input to different types of model architectures\cite{carl+19rmp}.
In many cases, this involves describing the spatial arrangement of point particles\cite{bron+21arxiv, widdowson2023recognizing} -- e.g. atoms in the construction of microscopic models of matter -- in a way that is consistent with the basic physical symmetries governing the relationship between a structure and its properties\cite{thom+18arxiv,vill+21arxiv}.
It would be desirable to determine a \emph{complete} set of invariants, i.e. a set of geometric descriptors that are equal for two configurations if and only if these configurations are the same, modulo translations, rotations, inversion, and permutations of equivalent atoms (see also Sec. 5 of Ref.~\citenum{musi+21cr} for a discussion of different definitions of completeness).
Unfortunately, it was recently shown that many popular structural representations are \emph{incomplete},\cite{pozd+20prl,pozd+22mlst} leading to the outright impossibility of predicting different values for the properties of structures even though they are unrelated by symmetry, and more broadly to numerical instabilities that extend to nearby configurations \cite{pars+21mlst,pozd+21ore}.

Even though the significance of this problem for data-driven atomistic modeling has been recognized only recently,  the problem of describing an arrangement of particles based on limited and/or symmetrized information is a long-standing one \cite{patt44pr}. 
This issue is the subject of a substantial body of literature, touching upon open problems in invariant  theory\cite{bout-kemp04aam} and the Euclidean distance geometry problem\cite{libe+14siamr,dokm+15ieeesp,tasi-lai19ieeetit}, with applications to fields as diverse as signal processing and the determination of protein structure by NMR. 
The broad connection to geometric deep learning and graph convolution\cite{bron+21arxiv}, as well as to the construction of equivariant physical models\cite{vill+21arxiv} is also well-understood. 
In the application domain that we focus on here, symmetry has been widely used since the early efforts to construct machine-learning interatomic potentials\cite{behl-parr07prl,bart+10prl,rupp+12prl}, and it also underlies efforts geared towards recognizing the (dis)similarity between structures in simulations and databases\cite{sade+13jcp,zhu+16jcp}. 

A systematic study of the representations used to characterize atomic structures\cite{will+19jcp,musi+21cr} has revealed that many of these frameworks rely on the same physics-inspired descriptors to characterize groups/clusters of atoms relative to a central one: discretized versions of structural correlation functions of the neighbor density with increasing \emph{body order}\cite{seth06book}  - which are equivalent to the unordered list of interatomic distances, angles, dihedrals, relative to the central atom. 
While it was shown that keeping \emph{all} orders of correlations provides a complete, symmetry-adapted \emph{linear} basis to expand structure-property relations\cite{shap16mms,duss+22jcp}, the low-order versions -- that are most commonly adopted in practice -- cannot distinguish groups of \emph{degenerate} configurations \cite{pozd+20prl}.
Similar problems affect other machine-learning architectures, such as distance-based graph convolutional networks\cite{pozd+22mlst}, and even equivariant neural networks\cite{thom+18arxiv,ande+19nips}, that are, in theory, universal approximators, but rely on constructing a hierarchy of symmetric polynomials that are closely related to the density-correlation features discussed above\cite{niga+22jcp2,bata+22arxiv}. 
Completeness, in this context, is only guaranteed in the limit of infinitely deep networks, which makes it harder to allocate computational effort between different architecture parameters, such as network depth and basis resolution~\cite{duss+22jcp}.
In this Letter, we discuss a construction that relies only on correlations between a finite number of neighbors to define invariant descriptors of a local atomic environment, that are complete in the sense that they can be used to build simple universal approximators for atomic properties and to differentiate between arbitrary configurations while scaling favorably with respect to system size and allowing for efficient implementations.

\begin{figure}
\includegraphics[width=1.0\linewidth]{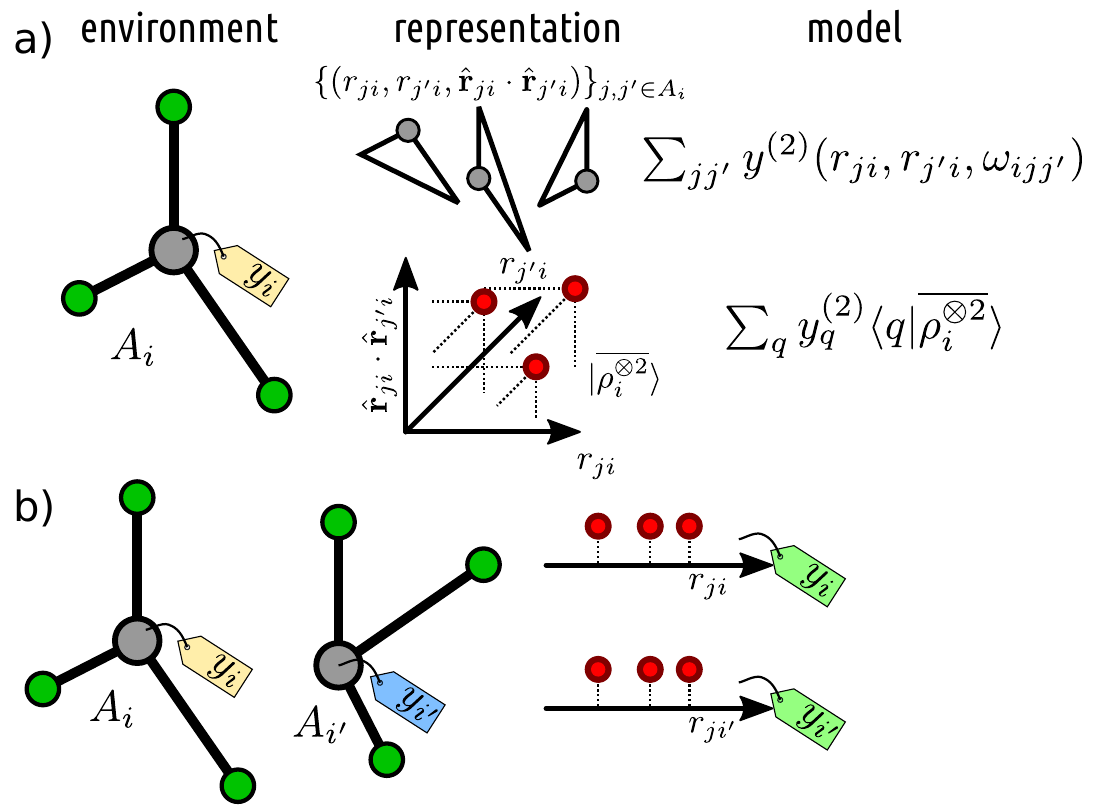}
\caption{
(a) Representations of an atomic environment $A_i$ in terms of neighbor-atom pairs, or symmetrized three-point correlations. The two schemes are equivalent in the complete basis set limit and generalize to higher-order representations. 
(b) A pair of environments that are degenerate to $\nu=1$ (list-of-distances) representations. 
\label{fig:scheme-rhoi}
}
\end{figure}

To appreciate the significance of this result without delving into the theory of density-based representations (interested readers may refer to the discussion in Ref.~\citenum{musi+21cr}, or to the summary in the SM) it is useful to consider the concrete case of two-neighbor representations  (Fig.~\ref{fig:scheme-rhoi}a).
To evaluate a property $\by(A)$ of an atomic structure $A$, it is convenient to express it as a sum of atom-centered contributions $\by(A_i)$, that in turn can be expanded into body-ordered terms $\by^{(\nu)}(A_i)$ depending simultaneously on the relative position of $\nu$ of the neighbors of the central atom $i$.
In line with previous literature, we indicate terms associated with correlations between $\nu$ neighbors as having body order $(\nu+1)$.
In the case of a scalar target (e.g. the interatomic potential) the two-neighbor term can be expressed as a sum of a function of two distances and one angle, $y^{(2)}(r_{i_1i}, r_{i_2 i}, \brhat_{i_1i}\cdot \brhat_{i_2i})$ over all neighbors pairs.  
Alternatively, and equivalently\cite{shap16mms,will+19jcp,drau19prb,duss+22jcp}, we can expand the histogram of $(r,r',\theta,)$ on a discrete basis $\rep<q|$, and approximate $y^{(2)}(A_i)$ as a linear combination of these coefficients, $\sum_q y_q^{(2)}\rep<q||\frho_i^2>$. 
Many widespread descriptors, from atom-centered symmetry functions\cite{behl11jcp} to the SOAP power spectrum\cite{bart+13prb} can be written in one of these forms. The resolution of the basis is a convergence parameter that in practical applications\cite{gosc+21mlst,duss+22jcp,bigi+22jcp}, and in particular when using the descriptors as inputs of a non-linear model\cite{bazt+22ncomm}, can be usually converged with ease. 

Higher-order terms can be obtained by computing $(\nu+1)$-body symmetric density correlations -- that we indicate with the shorthand $\rep|\frho_i^\nu>$. {This body-ordered expansion of $\by(A)$ converges systematically with $\nu$ but is rather cumbersome. 
Instead, many frameworks rely on low-$\nu$  descriptors and use them as input of artificial neural networks\cite{behl-parr07prl}. 
The non-linear functional form generates \emph{some}, but not all, the terms with higher body order\cite{glie+18prb,niga+22jcp2}, leading to more flexible models and making more effective use of the features restricted to a \emph{finite} body order $\nu$. 

Finite body-order descriptors, however, have a fundamental limitation: low-$\nu$ invariants are not \emph{complete}, in a way that cannot be remedied by non-linear transformations.} 
For example, if one uses only interatomic distances ($\rep|\frho_i^1>$), it is easy to realize multiple structures with neighbors placed in arbitrary directions but the same list of distances $\{r_{ji}\}$  (Fig.~\ref{fig:scheme-rhoi}b). All these structures will have identical $\rep|\frho_i^1>$ features, and therefore also the same output for any non-linear model built on them.
Less trivial constructions have been found that lead to degenerate pairs of configurations for triangle-histogram features (the power spectrum $\rep|\frho_i^2>$) as well as for tetrahedra-histogram features (the bispectrum $\rep|\frho_i^3>$)\cite{pozd+20prl}.
Counterexamples for higher $\nu$ are not known, but there is no proof of completeness either: for a general function no finite order $\nu$ of rotationally symmetric correlations suffices for a complete reconstruction\cite{kaka12jmiv}.
The existence of degeneracies means that, for finite $\nu$, there are pairs of distinct structures for which it is \emph{impossible} to build an ML model that predicts the differences in their properties, regardless of the resolution of the descriptors. 
It also determines  numerical instabilities in their vicinity, and therefore models built on these incomplete features can suffer from degraded accuracy and transferability\cite{pozd+21ore,pars-goed22jcp,pozd+22jcp}.

\begin{figure}[tbp]
\includegraphics[width=1.0\linewidth]{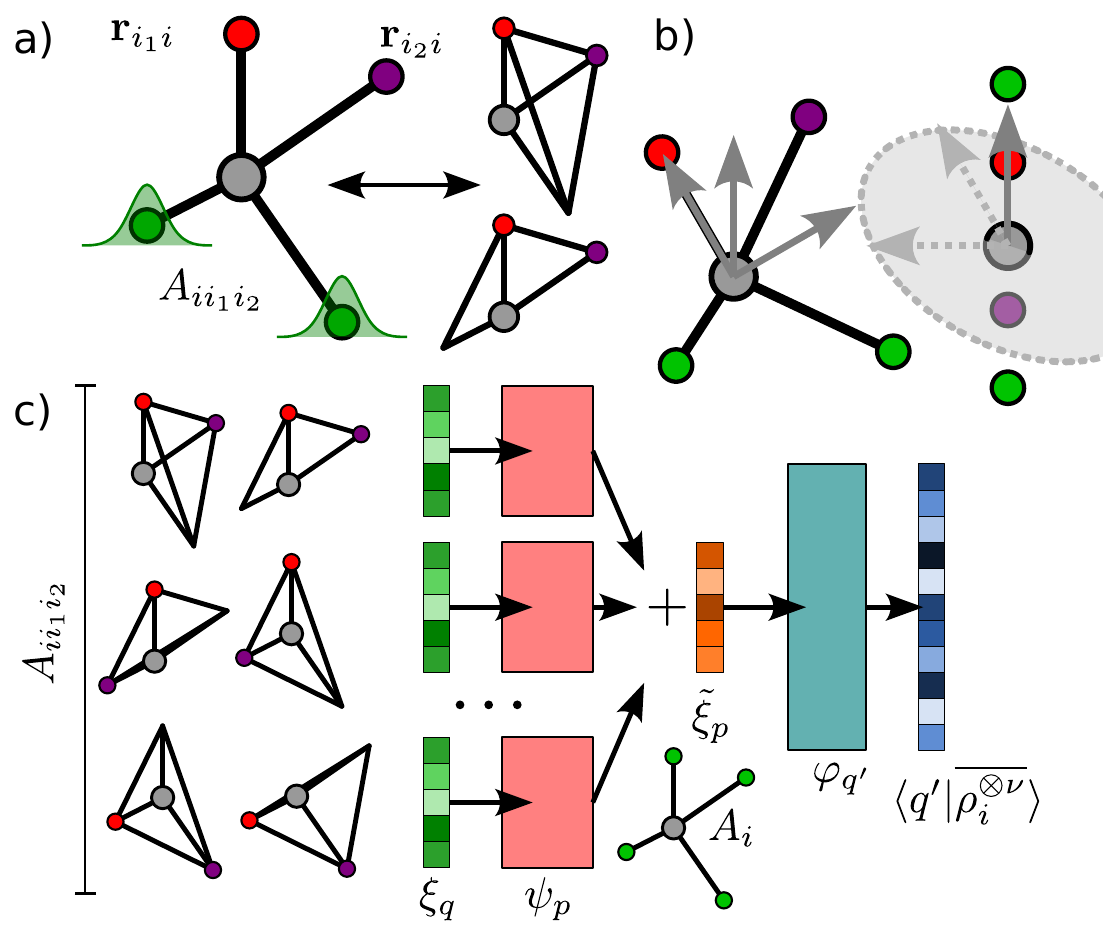}
\caption{
(a) 3-center-1-neighbor features describe the position of atoms relative to two tagged atoms around the center $i$. This is equivalent to a list of tetrahedra sharing the $ii_1i_2$ triangle.
(b) Tagging two atoms defines a local orthogonal coordinate system - except when the tagged atoms are collinear, and the azimuthal directions are ill-defined. 
(c) An encoder-decoder architecture can be used to determine non-linear \emph{environment} features $\tilde{\feat}_b$ that contain enough information to reproduce high-order representations.
\label{fig:scheme-nn}
}
\end{figure}

To avoid these fundamental and practical flaws we want to construct a framework {based on neighbor densities} that is provably complete while using a finite order of correlations between neighbors. 
To do so, we use an extension of the density-correlation framework to $(N+1)$-center quantities, i.e. descriptors of $N$-tuples of atoms in the environment $A_i$, that are invariant to the ordering of their neighbors, but equivariant to the labeling of the $N$ tagged particles\cite{niga+22jcp}.
The central idea in our construction is to build invariant descriptors representing the histogram of one neighbor ($\nu=1$) in the coordinate system defined by the central atom and two tagged neighbors ($N=2$), translating complete frameworks based on local coordinate systems\cite{zhan+18prl,puny2021frame,kurlin2022exactly,kurlin2022computable,pozdnyakov2023smooth} in the language of density correlation descriptors. 
Using Dirac's bra-ket notation to enumerate the components of the descriptor, we compute features of the following form: 
%
\begin{multline}
\vspace{-2mm} 
\rep<n_1 l_1; n_2 l_2; n_3 l_3||\frho_{{ii_1 i_2}}^1> \propto\!\!\!\!
\sum_{m_1 m_2 m_3} \!\!\!\!\cg{l_1 m_1}{l_2 m_2}{l_3 m_3}  \\
\times \rep<n_1 l_1 m_1||\br_{i_1 i}> \rep<n_2 l_2 m_2||\br_{i_2 i}>
\rep<n_3 l_3 m_3||\rho_i>. 
\label{eq:rho1-ii1i2}
\end{multline}
This expression involves an expansion $\rep<nlm||\br_{ji}> \equiv R_{nl}(r_{ji}) Y^m_l(\brhat_{ji})$, in radial functions $R_{nl}$ and spherical harmonics $Y^m_l$, of the position of neighbors $i_1$  and $i_2$.  $\rep<nlm||\rho_i> = \sum_j \rep<nlm||\br_{ji}>$ is the expansion of the neighbor density on the same basis. 
The contraction with Clebsch-Gordan coefficients $\cg{l_1 m_1}{l_2 m_2}{l_3 m_3}$ selects the rotationally invariant parts. 
The first ingredient in our approach is the realization that Eq.~\eqref{eq:rho1-ii1i2} provides a complete description of the environment $A_i$. 
A detailed proof is reported in the SM, but the reasoning can be explained in relatively simple terms. 
As long as atoms do not overlap, the peaks in the neighbor density precisely encode information on the position of all the atoms.
Averaging $\rep|\rho_i>$ directly over rotations discards angular information. 
However, the presence of descriptors of the position of two tagged neighbors $(i_1, i_2)$ means that rotational averaging of the entire tensor product preserves the full information on the position of all neighbors \emph{relative to these two atoms} (Fig.~\ref{fig:scheme-nn}a). 
In other terms, the descriptors in Eq.~\eqref{eq:rho1-ii1i2} are equivalent to knowledge of the position of the neighbors in the coordinate system defined by $(\br_{i_1 i }, \br_{i_2 i})$.
Provided that $\br_{i_1 i }$ and $ \br_{i_2 i}$ are not collinear, it is possible to reconstruct the position of all the atoms within the environment relative to the center $i$, modulo rotations.
Thus, by feeding a sufficiently fine-grained discretization of $\rep|\frho_{{ii_1 i_2}}^1>$ to a universal approximator, any function of the coordinates of the atoms in $A_i$ can be approximated in a way that is invariant to rotations, and to neighbor permutations, \emph{except for the tagged atoms} $i_1$ and $i_2$. {As we shall see, the convergence with the number of three-center descriptors and non-linear features can be achieved easily.}
In order to restore invariance and recover a description of the single-center environment, we sum over these two indices.
Performing this sum for Eq.~\eqref{eq:rho1-ii1i2} recovers the bispectrum $\rep|\frho_i^3>$ (recall that  $\sum_j \rep|\br_{ji}> = \rep|\rho_i>$) that we know to be incomplete\cite{pozd+20prl}.  
Instead, we apply a non-linear layer on top of the three-center ($N=2$) features \emph{before} summing over $i_1$ and $i_2$. 
Information on individual pair-dependent features is discarded, but any property of the environment $A_i$ can be learned, provided that one uses a sufficiently expressive architecture.  
Defining the shorthand $\feat_q(A_{ii_1i_2})= \rep<q||\frho_{{ii_1 i_2}}^1>$, where $q$ is a compound index that encompasses the labels $(n_1, l_1, n_2, l_2, n_3, l_3)$, 
we can write the environment features
\begin{equation}\label{eq:feat-psi-p}
\tilde{\feat}_p(A_i) = \sum_{i_1i_2} \psi_p[\bfeat(A_{ii_1i_2})].
\end{equation}
If one chooses $\psi_p$ to be a scalar universal approximator, Eq.~\eqref{eq:feat-psi-p} implements a \emph{deep set} architecture\cite{deepsets}, which is known to provide \emph{permutation invariant} universal approximation. 
Combining information on \emph{all} $(i_1,i_2)$ pairs within $A_i$ also ensures completeness in the presence of collinearity of certain pairs. 
For pairs corresponding to collinear vectors, $\rep|\frho_{{ii_1 i_2}}^1>$ cannot resolve the position of the neighbors along the azimuthal direction (Fig.~\ref{fig:scheme-nn}b). 
However if \emph{some} of the neighbors in $A_i$ are not collinear, there will be at least one pair that forms a basis for three dimensions and provides a complete description, and the compound features $\tilde{\bfeat}$ incorporate this information. 
If all the atoms are aligned along the same axis, then there is no azimuthal information to be encoded, and the description remains complete. 

It is important to note the relationships between this procedure and several ideas that have been used in the literature, built around the definition of a rigid coordinate system relative to which one can build local (or global) geometry descriptors \cite{bere+15jctc,zhan+18prl}, or to that of summing over all (or some) neighbor index permutations, a representation that is invariant to rotations but not to neighbor ordering \cite{mous12prl, hans+13jctc}. 
These approaches typically suffer from a lack of smoothness, because the procedure that determines the reference environment, or the index permutations that should be included in the summation, leads to discontinuous variations as a structure is deformed continuously.
The recently-introduced Weighted Matrices Invariant (WMI) framework\cite{kurlin2022computable,widdowson2022resolving} is the closest to the spirit of our construction.  It uses the unordered set of local atomic coordinates as an invariant and provides a practical algorithm to compare structures using a suitably defined smoothly-varying \emph{metric}. 
Our approach, instead, effectively defines body-order-complete, smooth, invariant \emph{descriptors} that can be readily understood and used in relation with several popular schemes based on discretized correlations of the neighbor density. A more in-depth comparison of the current approach with alternatives can be found in the SM.

The expression in Eq.~\eqref{eq:feat-psi-p} does \emph{in principle} have sufficient flexibility to fit arbitrary symmetric functions of the atoms within $A_i$. 
For example, a single universal approximator $\psi$ would suffice to make a complete predictor of a (short-ranged) interatomic potential, that would effectively be obtained as a sum of triplet energies. 
From this point of view, it is interesting to note the excellent performance that has been recently shown\cite{musa+23ncomm} for ML potentials that are built as a sum of \emph{pair} terms, which goes one step towards the provably-complete framework we propose here. 
It is easy to see, however, that a single feature cannot provide a complete description of an environment: a simple counting argument implies that at least $(3 n_i-6)$ features are needed to completely describe the degrees of freedom of an environment that contains $n_i$ atoms. 
We propose to use an encoder-decoder architecture to generate a set of non-linear features $\tilde{\bfeat}$ that can predict a large set of higher-order invariants (ideally with $\nu$ approaching the number of neighbors), as a way to generate a complete description (Figure~\ref{fig:scheme-nn}b). 

\begin{table}[tp]
    \centering
    \begin{tabular}{l r@{}c@{}l c c }
    \hline\hline
       \multicolumn{4}{c}{Model}    & \multicolumn{2}{c}{Relative Validation Error}   \\
  Name &  \multicolumn{3}{c}{Architecture} & Mean $A^\pm$ & Diff. $A^\pm$ \\
    \hline   
A$_\text{L}^{\rho}$  &  $\rho_{ii_1i_2}^{\otimes 1}$$\substack{\longrightarrow\\\text{L}}$ & $\tilde{\feat}_p$  & $\substack{\longrightarrow\\\text{NL}}$ $\rho_i^{\otimes 7}$        &   0.017   &   1.00
   \\
A$_\text{L}$  &  & & $\!\!\!\!\substack{\hookrightarrow\\\text{NL}}$ $E$  & 0.172 & 1.00  \\
A$_\text{NL}^{\rho}$  &  $\rho_{ii_1i_2}^{\otimes 1}$$\substack{\longrightarrow\\\text{NL}}$ & $\tilde{\feat}_p$  & $\substack{\longrightarrow\\\text{NL}}$ $\rho_i^{\otimes 7}$        &   0.006      &    0.143
   \\
A$_\text{NL}$  &  & & $\!\!\!\!\substack{\hookrightarrow\\\text{NL}}$ $E$  & 0.110 & 0.457 \\
B$_\text{L}$  &  \multicolumn{3}{c}{$\rho_{ii_1i_2}^{\otimes 1}$$\substack{\longrightarrow\\\text{L}}$  \!$E$\!  }  & 0.385    & 1.00 \\
B$_\text{NL}$  &  \multicolumn{3}{c}{$\rho_{ii_1i_2}^{\otimes 1}$$\substack{\longrightarrow\\\text{NL}}$  \!$E$\!  }  &  0.017  & 0.199 \\
C$_\text{L}$ &   \multicolumn{3}{c}{$\rho_{i}^{\otimes 7}$$\substack{\longrightarrow\\\text{L}}$  \!$E$\!  }  &  0.127   & 0.887 \\
C$_\text{NL}$ &   \multicolumn{3}{c}{$\rho_{i}^{\otimes 7}$$\substack{\longrightarrow\\\text{NL}}$  \!$E$\!  }  &  0.100 & 0.431 \\
D$_\text{L}$ &   \multicolumn{3}{c}{$\rho_{i}^{\otimes 3}$$\substack{\longrightarrow\\\text{L}}$  \!$E$\!  }  &  0.389 & 1.00 \\
D$_\text{NL}$ &   \multicolumn{3}{c}{$\rho_{i}^{\otimes 3}$$\substack{\longrightarrow\\\text{NL}}$  \!$E$\!  }  &  0.168 & 1.00 \\
    \hline\hline
    \end{tabular}    
\caption{Accuracy of different models (we indicate with `L' and `NL' linear and non-linear layers) for learning the high-$\nu$ features, and the energy of a validation set of 500 bispectrum-degenerate B$_8$ pairs. 
We separately report the error on the mean of the features and energies of degenerate pairs $(A^+,A^-)$, and on their difference, normalized by the standard deviation in the validation set -- i.e. for the error on the mean $\epsilon(\langle E_\pm \rangle )^2=\frac{1}{n_\text{test}} \sum_{A^\pm} |\frac{1}{2}(\tilde{E}(A^+)+\tilde{E}(A^-)) - \frac{1}{2}({E}(A^+)+{E}(A^-))|^2$
and the difference $\epsilon(\Delta E_\pm )^2=\frac{1}{n_\text{test}} \sum_{A^\pm} |(\tilde{E}(A^+)-\tilde{E}(A^-)) - ({E}(A^+)+{E}(A^-))|^2$.
We report the error relative to the standard deviations $\sigma(\frac{1}{2}({E}(A^+)+{E}(A^-))) $ and $\sigma({E}(A^+)-{E}(A^-))$, respectively. Absolute energy errors can be obtained by multiplying the values by 4.26 eV (Mean $A^\pm$) and 0.245 eV (Diff. $A^\pm$).    }
    \label{tab:results}
\end{table}

\begin{figure}[tbp]
\includegraphics[width=1.0\linewidth]{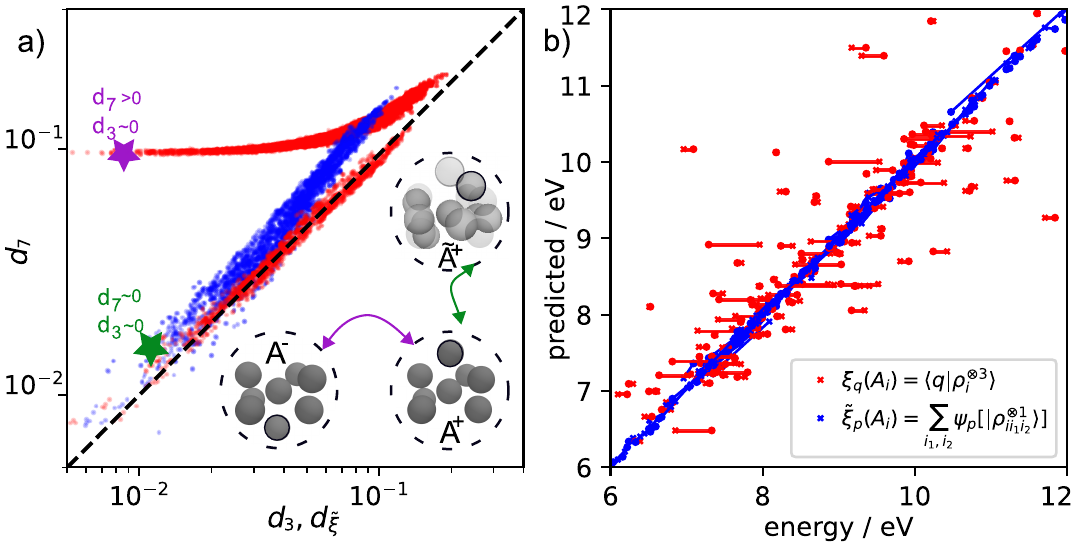}
\caption{(a) Correlation plot of the distances between pairs of structures from the B$_8$ dataset, computed based on the bispectrum ($d_3$, red), the intermediate encoded features  from Model A$^\rho_{\text{NL}}$ ($d_{\tilde{\xi}}$, blue), and the full-body features ($d_7$), taken as fully-discriminating descriptors. Configurations are randomly distorted ($\tilde{A}^+$ in the image), to reveal the behavior in the vicinity of the singular points (a bispectrum degenerate pair $A^+$ and $A^-$ is also shown in the inset), for which the bispectrum distance would be exactly zero. 
(b) Predictions for a representative 250 pairs   of the validation set from model B$_{\text{NL}}$(blue) and D$_{\text{NL}}$ (red). 
Degenerate pairs are joined by a line. The bispectrum-based predictions are identical for the pair, whereas the triplet features $\tilde{\xi}$ can resolve the difference in energy of the pair and achieve better accuracy overall.
}
\label{fig:boron-energy}
\end{figure}

To demonstrate this idea, we take the family of degenerate pairs of environments from Ref.~\citenum{pozd+20prl} that are indistinguishable with respect to all atom-centered invariants up to the $\nu=3$ bispectrum. 
This construction involves 7 neighbors around a central atom and generates a manifold with six continuous geometric parameters, and can be decorated with up to three atom types. For simplicity, we realize it with a single element, choosing boron clusters B$_8$ that are known to exhibit remarkable structural diversity\cite{bous97prb,de+11prl}.
By varying the parameters in an appropriate range, we generate a dataset of 4000 pairs, computing their energy with PBE-DFT in the cc-pVDZ basis, using PySCF\cite{sun+20jcp}  
(details and resulting structures are given in the SM). 
We then compute both 3-center-1-neighbor descriptors of $A_{ii_1 i_2}$ triplets, as well as atom-centered invariant correlation features of order $\nu=7$, based on the NICE framework\cite{niga+20jcp}.
We emphasize that - given that the degeneracy affects the \emph{environment} and is resolved by including multiple centers - all descriptors and models are built for a single atom at the origin. 
We build two simple encoder/decoder architectures that combine 3-center descriptors, contract them into environment descriptors, and then apply a multi-layer perceptron (MLP) with SiLU activation to approximate the high-order features.
One architecture (A$_\text{L}$ model) uses only a linear transformation before summing over $(i_1, i_2)$, while the other (A$_\text{NL}$ model) also includes a non-linear layer (another MLP) \emph{before} the contraction. Details of the architectures are given in the SM, but are not essential to our findings. 
We use the intermediate $\tilde{\bfeat}(A_i)$ features in the two models to learn the energies of the B$_8$ clusters. 
We also build ``direct'' models for the energy as a sum of linear and non-linear triplet energies (B$_\text{L}$ and B$_\text{NL}$), full-body-ordered atom-centered features (C$_\text{L}$ and C$_\text{NL}$), and bispectrum features (D$_\text{L}$ and D$_\text{NL}$).
Table~\ref{tab:results} summarizes the validation set accuracy of all these architectures. 
The accuracy in reproducing the full-body-order $\rep|\frho_i^7>$ correlation features can be interpreted as a measure of the resolving power of features, while the accuracy in predicting energies measures the impact of lack of resolution on a common regression task. 

As expected, all models that rely on a linear contraction of the 3-center descriptors $\rep|\frho_{{ii_1i_2}}^1>$, being equivalent to models based on bispectrum, are incapable of resolving the degenerate pairs, leading to a fractional error of 1 on the difference of features and energies. 
The lack of resolving power, and the consequent loss of regression accuracy, are also evident in the sensitivity analysis in (Fig.~\ref{fig:boron-energy}a), and in the energy prediction parity plot (Fig.~\ref{fig:boron-energy}b).
Even though the performance of individual models depends in a non-trivial way on the model details and the train-set composition and size, one can make a few observations based on the fact that we kept a consistent architecture for the non-linear layers among all models.
For starters, the relative errors on feature and energy differences among degenerate pairs are large also for the models that have -- in principle -- universal approximating power. 
This is due in part to the fact that these differences are small compared to the overall variation in energy, and that the magnitude of correlation features is dominated by low-order terms. 
Furthermore, the practical model accuracy depends on the fine-tuning of the architecture, which we deliberately avoided to facilitate a qualitative comparison between models. 
For example, one can clearly see that the NICE discretization of $\rep|\frho_i^7>$ is not sufficiently converged to provide a complete \emph{linear} basis: the C$_\text{L}$ model has a validation set error of almost 90\%{} for the energy difference between pairs. 
As for many frameworks in the literature, this lack of basis set convergence can be mitigated by applying a non-linear layer to the high-order features: both the C$_\text{NL}$ and the A$_\text{NL}$ models (the latter being computed on top of the latent features that allow the reconstruction of $\rep|\frho_i^7>$ to a very good accuracy) achieve an error on the energy differences around 45\%. 
From this point of view, the fact that a latent layer with only 128 features performs similarly to the full set of $\rep|\frho_i^7>$ (that is iteratively contracted to hold $\approx$8000 features) {underscores the effectivenes of non-linearities in tackling issues of basis set convergence.} 
The error in approximating  $\rep|\frho_i^7>$ converges quickly with latent-space size (see SM), which is consistent with the notion that a minimal set of symmetric descriptors does not have to be much larger than the number of degrees of freedom in the atomic environment being described, even though we cannot provide a rigorous upper bound to the number of non-linear features needed to be complete.  
Finally, we observe that the B$_\text{NL}$ model based \emph{directly} on non-linear functions of the 3-center features performs much better than all other options -- not only in terms of accuracy in describing pair differences, but also the mean energy of each pair.
The impact of degeneracies on realistic data sets, that do not explicitly include them, depends on the details of the model\cite{pozd+22jcp}, and previous studies of power spectrum degeneracies showed that a clear effect is only observed in the limit of very large datasets\cite{pozd+20prl}.


Summarizing our findings, we propose a practical solution to the open problem of determining a complete local representation of an atomic structure (more generally, a 3D point-cloud) based on a finite-dimensional and low body-order $O(3)$-invariant description of an environment. 
We demonstrate the idea on an artificial, but very challenging, set of \ce{B_8} cluster configurations, designed to defy linear atom-centered descriptors of the same body order. 
The accuracy and computational cost of an ML model in real-life scenarios depend on many different considerations. 
This approach indicates a way forward to designing ML frameworks that have universal approximation power while being smooth, symmetric, and free from the restrictions of architectures based on $O(3)$ equivariant layers (see e.g. Ref.~\citenum{pozdnyakov2023smooth} for a very recent example of a model that achieves excellent performance, enforcing equivariance by averaging over an ensemble of coordinate systems).
The close connection with established, well-understood body-ordered descriptors means that the ideas we expose here can be easily incorporated into existing frameworks, such as those based on atom-centered symmetry functions or those that build non-linear models based on invariant density-correlation features, which can now be safely truncated to a fixed and low order by evaluating features and models at the level of neighbor triplets.

\begin{acknowledgments}
MC and JN acknowledge funding from the European Research Council (ERC) under the research and innovation programme(Grant Agreement No. 101001890-FIAMMA) and the NCCR
MARVEL, funded by the Swiss National Science Foundation (SNSF, grant 
number 182892). MC, SP and KKHD acknowledge support from the Swiss Platform for Advanced Scientific Computing (PASC).
\end{acknowledgments}

\section*{Supporting Information}
The supporting information contains a detailed derivation of the main result and a discussion of the relation to alternative approaches to build symmetry-invariant descriptors, as  well as a description of the architecture of the models we used in our example. The dataset used for benchmarks, as well as code to train and evaluate the different models are available at Ref.~\citenum{zenodo-three-center}

\section*{Author Declarations}
The authors have no conflicts of interest to declare.

\onecolumngrid
\newpage
\renewcommand{\thefigure}{S\arabic{figure}}
\renewcommand{\theequation}{S\arabic{equation}}
\renewcommand{\thetable}{S\arabic{table}}

\section{Summary of density-based descriptors}
\label{sec:reps}
A structure $A$ containing $\natoms$ independent atoms with positions $\{ \br_i \}$  and chemical types $\{ a_i \}$ can be described within the framework of atom-centered density correlations (ACDC) as follows. 
As the name suggests, this family of descriptors or representations decomposes the structure into \emph{environments} $A_i = \{ a_i, \{(a_j, \br_{ji}):r_{ji} < r_\text{cut}\} \} $ centered on each atom, consisting of neighboring atoms (expressed by their inter-atomic distance vectors $\br_{ji}=\br_j-\br_i$ and species $a_j$) within a certain distance $r_\text{cut}$ away from the center. Usually, the neighbors or equivalently their separation from the center $\br_{ji}$ is captured through a localized function $g(x-\br_{ji})$ that peaks at the position of the neighbor, for example, $g$ could be a Gaussian function or a Dirac-delta distribution (when this is the case we use the notation $g\rightarrow \delta$). 

This description of environments in terms of interatomic displacements is invariant to the choice of an arbitrary origin (equivalently, invariant to rigid translations). By invariance to a transformation, we mean that the description can be written as a function that remains unchanged when the transformation is applied to the arguments/variables. Historically, these descriptions are then made independent of the \emph{order} in which the neighbors are enumerated. This is easily achieved by an ``invariant pooling", e.g. a summation over all neighbors in the environments $A_i$, resulting in a translation and permutation invariant atomic neighbor density $\rho_i(\bx)$.
In the following discussion, we will use Dirac's bra-ket notation to capture a broad class of descriptions (such as SOAP, ACE, FCHL, etc.)
following a similar procedure, and attribute the specificities of each representation to the choice of basis and the localized function used to enumerate the neighboring atoms. So $g(\bx-\br_{ji}) \equiv \rep<\bx||\br_{ji}>$, and $\rho_i(\bx) \equiv \rep<\bx||\rho_{i}>$. 
It is easy to see that in the $g\rightarrow \delta$ limit this representation provides a \emph{linear} basis to expand properties that depend separately on the position of individual neighbors, 
\begin{equation}
\label{eq:neigh-contrib}
y^{(1)}(A_i)=\int \D{\bx} y^{(1)}(\bx) \rho_i(\bx) \underset{g\rightarrow\delta}{=}  \int \D{\bx} y^{(1)}(\bx) \sum_{j \in A_i} \delta (\bx - \br_{ji}) =
\sum_{j\in A_i} y^{(1)}(\br_{ji})
\end{equation}

Notice that the description of each atomic environment still depends on the orientation of the reference system chosen to describe the coordinates $\br_{ji}$ of the neighbor $j$ as the translational symmetrization only fixed the origin to the central atom. However, one might be interested in modeling scalar properties such as energies that are invariant to global rotations of the structure, or alternatively target an equivariant property such as the dipole moments that have a specific rotational behavior. To capture the behavior of the target with respect to rotations (or equivalently any other symmetry operation) of the structure, the structural descriptions can be endowed with the very same behavior by integrating over the corresponding symmetry operations. 

For instance, in the case of rotational equivariance, to integrate the neighbor density over rotations, it is easier to first write it on a basis of spherical harmonics $(\rep|lm>)$ that have a precisely determined rotational dependence.  
This change of basis from $\rep<\bx|$ to $\rep<nlm|$, where $\rep<n|$ is a discretized description of the absolute distance $|\bx|$ and $\rep<lm|$ captures the direction $\hat{x}$, can be written as
\begin{equation}
 \rep<nlm||\br_{ji}> \equiv \int \D{\bx} \rep<\bx||\br_{ji}> \rep<nl||x> \rep<lm||\bxhat>
\end{equation}
so that $\rep<nlm||\rho_{i}>= \sum_{j\in A_i }\rep<nlm||\br_{ji}>$ 

A representation with the desired symmetry (of rotation like the spherical harmonic $\rep|\glm>$) is then achieved by 
\begin{equation}
\rep<n||\frho[\glm]_i^1> \equiv\rep<n\glm||\rho_{i}> 
\end{equation}
in which we emphasize the SO(3) character of the representation by indicating $\glm$ in the ket.
Following this procedure, we have obtained a symmetric (equivariant) description of the target property in terms of contributions of individual neighbors. Eq.~ \ref{eq:neigh-contrib} still holds in the symmetrized form with $\rep<n||\frho[\glm]_i^1>$, providing a linear basis to expand a function with the symmetry of $Y_\lambda^\mu$ that depends on the coordinates of individual neighbors.

This procedure can be easily extended to \emph{body-ordered} expansions of atom-centered properties\cite{shap16mms,glie+18prb,will+19jcp,drau19prb}, that use the tensor products of $\rep|\rho_i>$ as a linear basis to express quantities that simultaneously depend on the position of $\nu$ neighbors. We use the shorthand $\rep<q||\frho[\glm]_i^\nu>$ to indicate such a description that is also equivariant. 

For more details on these atom-centered descriptions, we advise the reader to consult Refs. ~\citenum{musi+21cr, will+19jcp}. 
Notice that in constructing $\rep|\rho_i>$, we summed over all the neighbors in the environment, thereby losing the identity of an individual atom in the neighborhood, but retaining the identity of the single central atom $i$. An equivalent way of saying this is that $\rep|\rho_i>$ is a histogram of neighbor positions in  $A_i$ as it describes the distribution of the neighbor displacements without their explicit identities.
Similarly, for $\rep|\frho[\glm]_i^\nu>$, we started with a description of individual atoms that was then symmetrized over translations, permutations, and finally over rotations. This representation describes the histogram of correlations of the central atom with $\nu$ neighbors.

 One might be interested in descriptions of atomic environments retaining the identity of more than one atom, such as pairs, triplets or N-tuples of atoms. We proposed an extension \cite{niga+22jcp} of the ACDC formalism reviewed above to tag $N$ other particles to obtain a representation which explicitly addresses the identity of each of these atoms $\rep|\frho[\glm]_{{i i_1 i_2 \ldots i_N}}^\nu>$. 
 
 \begin{equation}
   \rep|\frho[\glm]_{{ii_1\ldots i_N}}^\nu > \equiv
\int_{SO(3)} \D{\Rhat}  \prod_{\alpha=2}^N \Rhat\rep|\br_{i_\alpha i}; g>  \prod_{\beta=1}^\nu \Rhat\rep|\rho_i> \otimes \Rhat \rep|\lambda\mu>
 \end{equation}
 These N-centered representations are translationally invariant and are also symmetrized in the same order- they are invariant to the permutation of the $\nu$ neighbors but \emph{equivariant} to the permutation of the labelling of the $N$-centers; and are rotationally symmetrized the same way as their one-center counterpart. We direct the interested reader to see the accompanying SI of Ref.~\citenum{niga+22jcp} for a detailed derivation. 
Note that with an $N$-centered description, we effectively describe the correlations of 1 central atom with $N$-1 other centers and $\nu$ neighbors. 
One can restore the N-centered representation to an atom-centered one by summing over the N-1 additional centers
\begin{equation}
    \sum_{i_1, \ldots i_N} \rep|\frho[\glm]_{{ii_1 i_2 \ldots i_N}}^\nu> = \rep|\frho[\glm]_{i}^{{\nu+N}}> 
\end{equation}
This provides a way one can promote neighbors to special centers or demote special centers to neighbors. Following the discussion above, a three-centered 1 neighbor representation is comparable to an atom-centered bispectrum, both of which describe correlations of one central atom with 3 other atoms, thus having similar descriptive power in terms of the correlation order captured. However, the former singles out indices of two of the neighbors while the latter is permutation invariant to all of the indices but the center $i$. 

\section{Proof of Completeness}

In this section, we provide a proof of the central result of the main text, namely that any continuous function $f(\br_{i_1i},\br_{i_2i},\dots,\br_{i_Ni})$ defined within the local neighborhood of a center atom $i$ that is invariant under rotations and permutations can be approximated to arbitrary accuracy using the features shown in Eq.~\eqref{eq:feat-psi-p}, 
\begin{equation}\label{eq:feat-psi-p-si}
\begin{split}
\tilde{\feat}_p(A_i) =& \sum_{i_1i_2} \psi_p[\bfeat(A_{ii_1i_2})], \\
\feat_q(A_{ii_1i_2})= &\rep<q\equiv (n_1, l_1, n_2, l_2, n_3, l_3)||\frho_{{ii_1 i_2}}^1>\\
\rep<n_1 l_1; n_2 l_2; n_3 l_3||\frho_{{ii_1 i_2}}^1> \propto&\!\!\!\!
\sum_{m_1 m_2 m_3} \!\!\!\!\cg{l_1 m_1}{l_2 m_2}{l_3 m_3} \rep<n_1 l_1 m_1||\br_{i_1 i}> \rep<n_2 l_2 m_2||\br_{i_2 i}>
\rep<n_3 l_3 m_3||\rho_i>. 
\end{split}
\end{equation}
The 3-vectors  $(\br_{i_1i},\br_{i_2i},\dots,\br_{i_Ni})$ indicate the positions of $N$ particles in $\bar{B}^3_{r_\mathrm{cut}} = \lbrace \br \in \mathbb{R}^3: \|\br\|\leq r_\mathrm{cut}\rbrace$ relative to some center atom $i$ ($\br_{ji} = \br_j - \br_i$) with coordinates given in an arbitrary coordinate system, and $\rep<nlm||\square>$ indicates discretization on a basis of radial functions and spherical harmonics.
The proof rests on two key steps: (1) using relative coordinate frames for a complete rotationally invariant description and (2) constructing a permutationally invariant function using these inputs. 
We devote subsections \ref{sec:proof_rotation} and \ref{sec:proof_permutation} to the two steps, and discuss in Section~\ref{sec:connections} the connections to other approaches.

\subsection{Complete Rotationally Invariant Descriptors using Relative Coordinate Frames}
\label{sec:proof_rotation}

For the first step, we will follow the presentation in Ref.~\citenum{kurlin2022computable} and use, for simplicity, the notation $(\br_{i_1 i},\dots,\br_{i_N i})$ to indicate the position of the neighbors relative to the central atom. Let us first assume that the first two particles at positions $\br_{i_1i}$ and $\br_{i_2i}$ are not collinear. These then define a new orthonormal basis $(\bv_1,\bv_2,\bv_3)$ of $\mathbb{R}^3$ by taking the first two basis vectors from the Gram-Schmidt orthonormalization procedure, i.e. by setting $\bv_1 = \frac{\br_{i_1i}}{\| \br_{i_1i} \|}$ and the second basis vector $\bv_2$ to be the normalized version of $\br_{i_2i} - (\br_{i_2i} \cdot \bv_1) \bv_1$, so that it lies in the plane spanned by $\br_{i_1i}$ and $\br_{i_2i}$. Finally, we set $\bv_3 = \bv_1 \times \bv_2$ leading to a unique right-handed coordinate system. 
The coordinates of an arbitrary point $\br_{ji}$ with respect to this relative basis are simply given by the inner products $\br_{ji} \cdot \bv_i$. In particular, these coordinates are invariant under rotations $\hat{R} \in \mathrm{SO}(3)$ as this reference frame rotates with the underlying structure. (A similar procedure is adopted in Ref.~\citenum{wang2018deepmd} for describing atomic environments.)\\
Apart from the non-collinearity assumption, which will be addressed shortly, we can repeat this construction by choosing any pair $(\br_{i_1 i}, \br_{i_2 i})$ to form the basis vectors from, leading to $N(N-1)$ rotationally invariant ways to describe the position of all the particles, which (after considering permutation invariance) maps to the weighted matrix invariants (WMIs) that have been proven to provide a metric that is able to differentiate between any pair of atomic structures \cite{kurlin2022computable}.

We now show that this is precisely the information contained in the 3-center descriptors in the main text. Consider the function (strictly speaking, a tempered distribution)
\begin{equation}
    f(\bx,\bx',\bx'') = \delta(\bx-\br_{i_1i})\delta(\bx'-\br_{i_2i})\delta(\bx''-\br_{i_3i}),
\end{equation}
which peaks whenever $(\bx,\bx',\bx'')$ are arranged in the same way as $(\br_{i_1i}, \br_{i_2i}, \br_{i_3i})$. By averaging over all rotations, we obtain precisely the 3-center feature for the special case of infinitely sharp $\delta$-like densities (i.e. $g\rightarrow \delta$ limit of atomic densities)
\begin{equation}
\!\!\!\!\rep<\bx;\bx';\bx''||\frho_{{ii_1i_2i_3}}^0> = \!\!\int\!\!\D{\hat{R}} 
\delta(\bx-\hat{R}\br_{i_1i})
\delta(\bx'-\hat{R}\br_{i_2i})
\delta(\bx''-\hat{R}\br_{i_3i}). \label{eq:n3-nu0-delta}
\end{equation}
There is a single peak whenever $(\bx,\bx',\bx'')$ are arranged as $(\br_{i_1i},\br_{i_2i},\br_{i_3i})$, up to a rotation. In particular, $\rep|\frho_{{ii_1i_2i_3}}^0>$ determines the position of $\br_{i_3i}$ relative to $\br_{i_2i}$ and $\br_{i_1i}$, just as the basis we have constructed before.\\
Summing over the third index $i_3$, we obtain
\begin{equation}
\rep|\frho_{{ii_1 i_2}}^1> = 
\sum_{i_3} \rep|\frho_{{ii_1i_2i_3}}^0>.
\end{equation}
This is essentially a histogram of the relative position of all the neighbors, that has (provided the atomic density peaks are sufficiently sharp, and the resolution fine enough) complete information on the position of all atoms within the environment $A_i$ relative to $i_2$ and $i_1$. 
A projection onto a \emph{complete} discrete basis preserves this information, which proves the completeness of $\rep<n_1l_1; n_2l_2; n_3l_3||\frho_{{ii_1 i_2}}^1>$. 
The size of this basis is a further convergence parameter, that is common to all atom-density correlation descriptors.

To fill in a final gap, we briefly address issues related to collinearity of the particles. Since we consider all $(\br_{i_1 i},\br_{i_2 i})$ pairs in the environment around $i$ ($A_i$), if only some of the particles are collinear but $A_i$ still contains at least one non-collinear pair $(\br_{i_1 i},\br_{i_2 i})$, the coordinate frame relative to these two atoms provides a complete description of the structure. A more formal proof that at least one such pair suffices can be found in Ref.~\citenum{kurlin2022computable}.
On the other hand, if all atoms in an environment are collinear, the environment can be completely reconstructed from the set of pairwise distances. 
This information is still contained in the descriptor $\rep|\frho_{{ii_1 i_2}}^1>$ due to the location of the peaks so that the 3-center descriptor allows reconstruction of the structure up to rotations.

\subsection{Permutation Invariance}
\label{sec:proof_permutation}

Recalling that our goal is to have a general expression for a function $f(\br_{i_1i},\br_{i_2i},\dots,\br_{i_Ni})$ that is invariant under rotations and permutations, we still need to provide the missing link from the rotationally but not yet permutationally invariant representation $\rep|\frho_{{ii_1 i_2}}^1>$. The crucial step is to perform a transformation that symmetrizes over the $N(N-1)$ possible choices of the two vectors $i_1$ and $i_2$ that form the reference basis.

For the weighted matrix invariants (WMIs) in Ref.~\citenum{kurlin2022computable}, the permutation invariance is taken into account by simply considering the $N(N-1)$ coordinate systems to be \textit{unordered} when comparing two structures.
In practical terms, checking whether the ordered tuple $(x_1,\dots,x_N)$ is equal to $(y_1,\dots,y_N)$ requires $N$ comparisons at most while checking whether two unordered sets $\lbrace x_1,\dots,x_N \rbrace$ and $\lbrace y_1,\dots,y_N \rbrace$ are equal naively suffers from an additional cost of checking through $N!$ ways to rearrange the elements. While Ref.~\citenum{kurlin2022computable} proposed more efficient methods to test the equality of two unordered sets, the methods are not directly translatable to common problems in condensed matter physics and related fields, in which we want to specify the descriptors for one structure rather than to compare two configurations.

At the same time, properties of functions invariant under permutations of the input variables have been studied in a more general context \cite{deepsets}. In particular, rephrasing the results for the case of three center representations, it has been shown (theorem 9 in Ref.~\citenum{deepsets}) that any permutationally invariant continuous function $f_\mathrm{perm}$ that takes a discretization  of the three-center features $\feat_q(A_{ii_1i_2})=\rep<q||\frho_{{ii_1 i_2}}^1>$ as inputs can be approximated arbitrarily well by functions of the form
\begin{align}\label{permutation_invariant_generic}
f_\mathrm{perm}\left(\bfeat(A_{ii_1i_2})\right) = F\left(\tilde{\bfeat}(A_i)\right),\quad\quad \tilde{\feat}_p(A_i) = \sum_{i_1,i_2}\psi_p(\bfeat(A_{ii_1i_2}))
\end{align}
where $F$ now is an arbitrary continuous function, and the $\psi_p$ are (non-linear) maps that transform the initial features, which are then summed over the indices $i_1$ and $i_2$ to generate the permutationally invariant intermediate features $\tilde{\bfeat} = (\tilde{\feat}_1,\dots,\tilde{\feat}_p,\dots)$. These precisely correspond to the components of the encoded intermediate features in Eq. \eqref{eq:feat-psi-p} in the main text. This completes the proof that any function $f(A_i)$ that depends on an atomic environment can be approximated arbitrarily well by the presented approach.

\section{Connection to related results}\label{sec:connections}

In this section, we summarize the connection between the three center descriptors proposed in this work and alternative viewpoints on the problem of achieving a complete description of an atomic environment. 

\subsection{Connection to the Gram matrix and Weighted Matrix Invariants}\label{connection_gram}

Let us denote the $N$ displacement vectors by $(\br_{i_1 i},\dots,\br_{i_N i})$.
If we only require descriptors that are invariant under the action of some improper rotation $\hat{R} \in \mathrm{O}(d)$, the $N\times N$ Gram-matrix defined by
\begin{align}
    G_{\alpha \beta} = \br_{i_\alpha i} \cdot \br_{i_\beta i}, \forall \alpha,\beta \in \lbrace 1,\dots,N\rbrace
\end{align}
is well known to be an overcomplete set of invariants  (see e.g. Ref.~\citenum{vill+21arxiv} for how this classical result can be used in an ML context). 

In fact, for $d=3$, if the first three vectors $\br_{i_1 i},\br_{i_2 i},\br_{i_3 i}$ are not coplanar, the first three rows (or equivalently, columns) of the Gram-matrix suffice, hence the overcompleteness.
While we need three (non-coplanar) rather than two (non-collinear) vectors here, this is consistent with the previous observation that we can reconstruct the entire structure from the relative coordinates of only $\br_{i_1 i}$ and $\br_{i_2 i}$. This is because the third of the three basis vectors $(\bv_1,\bv_2,\bv_3)$ mentioned in section \ref{sec:proof_rotation} is constructed using the first two. 
Thus, being able to take the cross-product to upgrade the vectors to a basis is what makes two vectors sufficient for a complete description of an environment.
For permutational invariance, the purely theoretical approach would be to simply consider the Gram-matrix up to permutation of the rows and columns, which is closely related to the WMIs in \cite{kurlin2022computable} and other similar constructions \cite{widdowson2022resolving}, that however avoid enumerating all permutations. Regardless of whether one uses inner products (Gram matrix) or the relative coordinates (WMIs) between the atomic positions, these sets of invariants are quite different from many of the more commonly used descriptors, in the sense that one is working with unordered sets. 
While defining a function on an unordered set is simple mathematically, a practical algorithm will in many cases require the storage of the elements in an ordered manner. Explicitly parametrizing a function on an unordered set would then require an additional specification of how to enforce the unordered nature. One such example is the summation over all permutations of indices, as is done with the invariants in this work. Thus, one cannot just use these “invariants” as naively as many of the other “descriptors” and, for instance, feed them directly into a neural network. These invariants are nevertheless useful to perform comparisons between structures due to the existence of a metric \cite{kurlin2022computable, widdowson2022resolving}.

A further issue with using invariants such as the unordered set of Gram-matrices or the WMIs is the fact that the number of features changes with the number of neighbor atoms. In principle, the two aforementioned methods could either be used to describe the structure as a whole, leading to a computational cost scaling as $O(N^2)$ in the number of atoms $N$, or be restricted to a finite environment using a cutoff radius. The second approach would be more favorable in terms of scaling, which then however introduces an extra complication due to discontinuities associated with atoms leaving or entering the local environment. In this case, not only the invariant matrices per se, but also their shape and number would change discontinuously.
 
As mentioned in the main text, similar approaches have been used to introduce approximate permutation invariance for models based on the distance matrix or its non-linear transformations.\cite{mous12prl, hans+13jctc}
This idea is also related to approaches, such as Social Permutation Invariant coordinates\cite{piet-andr11prl} or the Overlap Matrix fingerprints\cite{sade+13jcp} that compute non-linear functions of the distance matrix and then build descriptors from the sorted eigenvalues of the transformed matrix. While there is no proof that these descriptors are complete, they are empirically capable of differentiating between the known degenerate structures~\cite{pars+21mlst}. 

\subsection{Connection to other ML schemes in atomistic modeling}\label{connection_mtp_ace}

Within the subfield of condensed matter physics and materials science, there are two other approaches providing proofs of complete approximation capability starting from invariant descriptors. The first such approach was the moment tensor potential (MTP) \cite{shap16mms} proposed in 2016, which uses the Cartesian coordinates of the particles, followed by the atomic cluster expansion (ACE) in 2019 \cite{duss+22jcp} which can be seen as a related approach relying more on spherical coordinates.
We focus in particular on ACE and the mathematically equivalent methods including the NICE framework \cite{niga+20jcp} which are closer in spirit to our approach. All such models start by constructing an atom-centered neighbor density $\rep|\rho_i>$ around some atom $i$, either with Gaussians or infinitely sharp delta-functions, which by itself contains complete information about the atomic positions (for sufficiently sharp densities, the atomic positions are precisely the location of the peaks of the density function). As explained in section~\ref{sec:reps}, this representation is already permutationally invariant, but not yet invariant under rotations.
In practice, the densities are projected onto a basis, the coefficients of which $\langle nlm | \rho_i \rangle$ are used as inputs to further models. If a sufficiently large basis and sufficiently many coefficients of the expansion are retained, these still allow us to completely reconstruct the position of all atoms in the neighborhood.
Methods like ACE and NICE now use a linear model on all possible rotationally invariant combinations of these coefficients, which is mathematically equivalent to taking an arbitrary rotationally invariant (nonlinear) function of the inputs $f_\mathrm{rot}\left(\langle nlm | \rho_i \rangle\right)$, which can be written in a form similar to eq. \eqref{permutation_invariant_generic} using the shorthand $\zeta_q(A_i) = \rep< q\equiv (nlm) || \rho_i >$ as
\begin{align}\label{rotation_invariant_generic}
f_\mathrm{rot}\left(\boldsymbol{\zeta}(A_i)\right) = F( \tilde{\boldsymbol{\zeta}}(A_i)),  \tilde{\zeta}_p(A_i)  = \int_\mathrm{SO(3)}\mathrm{d}\hat{R}\, \varphi_p\left( \boldsymbol{\zeta}(\hat{R} A_i) \right),
\end{align}
where $F$ now is an arbitrary continuous function, and the $\varphi_p$ are (non-linear) maps that transform the initial features, which are then averaged over all possible rotations to generate the rotationally invariant intermediate features $\tilde{\boldsymbol{\zeta}} = (\tilde{\zeta}_1, \dots, \tilde{\zeta}_p, \dots, )$. In practice, these intermediate features would correspond to the ACE/NICE features that are used in the expansion.
In our approach, we start from a density $\rep|\frho_{{ii_1 i_2}}^1>$ that is invariant under rotations only, and then apply a generic permutationally invariant nonlinear function $f_\mathrm{perm}\left(\rep|\frho_{{ii_1 i_2}}^1>\right)$. At the purely conceptual level, apart from switching the orders between permutation and rotation symmetries, we can see that the two approaches are quite similar in spirit.

One key difference is that in many of the practical implementations including the original formulations of ACE and NICE, the body-order expansion is truncated at a finite order $\nu$. Thus, while complete in the limit of arbitrarily high body orders, the expansion is incomplete in practice. Applying a nonlinear model would technically generate higher order invariants, but in order to preserve a complete description, the model would still require that the atom-centered invariants are complete for some $\nu < N$, which is at present unproven.
From a mathematical point of view, this difference regarding implementations can be seen as a consequence of the fact that generating a generic permutationally invariant function, having the form shown in Eq.~\eqref{permutation_invariant_generic}, is significantly simpler than a generic function having rotational invariance. 
While we can apply arbitrary nonlinear functions and perform a sum over indices, rotationally invariant combinations are restricted to those obtained from a careful study of irreducible representations of the rotation group. The closest conceptual analogue to our approach would be to only take algebraically independent coefficients in the ACE/NICE expansion and use those as inputs into a nonlinear model.
The MTP approach essentially performs both steps at once, by constructing features that are invariant under both rotations and permutations from the get-go. More precisely, they are suitable rotationally invariant contractions of the atomic coordinates summed over all atoms in the structure, which can also be related to ideas of deep sets applied, with some modifications, to components of the Gram-matrix.

For all of the methods presented above, it should be noted that completeness for an arbitrary number of particles $N$ (excluding the center atom) is only guaranteed in a certain limit. For MTP and ACE/NICE, this is more directly seen as the maximal order of the used expansion, as well as (for ACE/NICE) the size of the basis used in the expansion.  
Even though a large number of features is usually necessary when these descriptors are used in a linear model, far fewer terms are necessary to achieve the notion of completeness that is used in this work, as is underscored by the aggressive truncation that can be used when introducing nonlinearities or message-passing operations in the architecture of related models e.g. in MACE\cite{bata+22nips}, NeQUIP\cite{bazt+22ncomm} or PAINN\cite{schu+21icml}.
For our invariants starting from the three center features, this truncation is hidden in the number of required encoded intermediate features $\xi_p$ as well as in the resolution of the discretization of the 3-centers-1-neighbor features.
Considering that $N$ atoms in an environment have $3N-3$ degrees of freedom modulo rotations, any complete model would need at least as many coefficients. In practice, apart from simple enough group actions, no simple bound for the minimal number of such features required to keep the representation complete is known. Some numerical experiments on the convergence of model performance as a function of the number of intermediate encoded features are shown in Fig.~\ref{fig:latent-space-error}.

In parallel, some of the authors in this work are also pursuing an alternative way to use the local coordinate systems defined by the triplets. Similar to the WMIs \cite{kurlin2022computable}, one makes use of the coordinates of the atoms with respect to this local frame. Instead of using unordered sets however, the coordinates are used more directly, albeit with some modifications, as inputs to a neural network for the final target property, which is discussed in a separate article \cite{pozdnyakov2023smooth}.

\section{Dataset generation and model details}
To generate a pair of bispectrum-degenerate structures we follow the construction in Ref.~\citenum{pozd+20prl}. We begin by placing an atom at the origin. A ring of radius $r$ with $n$ atoms (one oriented parallel to the $x$ axis, the others having azimuthal angles $\phi_i i\in \{1, \ldots n-1\}$) can then be placed with center (0,0,$+z_1$) and subsequently, the same ring with an azimuthal phase shift of $\psi$ is placed at a height of $-z_1$. An additional atom is then placed at coordinates (0,0, $\pm z_2$) to generate the $A^\pm$ structures.
From here we can see that the manifold of bispectrum-degenerate structures composed of $2n+2$ atoms is parameterized by $\{r, \phi_1, \ldots \phi_{n-1}, \psi, z_1, z_2 \}$. Additionally, the species of the central atom, the ring of atoms at $z=\pm z_1$, and the atom at (0,0,$\pm z_2$) can be different. 

The dataset of 4000 pairs of bispectrum-degenerate \ce{B_8} structures was generated by fixing $n=3$ (yet still being able to randomly position the atoms on the ring) and sampling these parameters randomly within a reasonable range (essentially selecting structures with minimal \ce{B-B} distances in a range comparable to that found in metastable cluster geometries). 
The energies for these structures were obtained by PBE-DFT calculations performed using PySCF in the cc-pVDZ basis. As an atom-centered descriptor of correlation order $\nu=7$ can in principle simultaneously describe all atoms in the structure, this representation is complete (in the limit of a complete basis). In the following, we compare the proposed three-center one-neighbor representation, which is universally complete, with the ACDC feature that is complete for this dataset. As a reminder, by completeness, we mean that the features take on unique values for structures unrelated by symmetric transformations that are already encoded in their construction (here, permutation of identical atoms, rigid translations and (im)proper rotations.

\subsection{Feature computation}
ACDC features up to $\nu=7$ and the three-center, one neighbor correlation ($N=2$, $\nu=1$) features were computed for 4000 pairs of degenerate \ce{B_8} structures with a cutoff radius of 2.0 \AA{} (that encompasses all atoms in the cluster), Gaussian smearing of $0.2$\AA{}  $n_\text{max}=6$, $l_\text{max}=3$. Finally, ACDCs from $\nu=1$ to up to $\nu=7$ were stacked together and compressed using PCA so that the final atom-centered descriptor had as many features as the number of structures.

\subsection{Details of the feature reconstruction models}

The encoder-decoder architectures for reconstructing the atom-centered feature (as described above, thus fixing the output size to the total number of structures ) from the three-centered description were implemented in PyTorch\cite{pytorch} and differed in the transformations employed in the encoder. Here, we describe two types of models, namely linearly encoded and nonlinearly encoded triplet features for a given latent space size (which we fix to be 128 for all models discussed in the main text). 
Linearly encoded models employed just a single linear layer to contract the input, whereas a nonlinear encoder sequentially contracted the input to the final latent size over three layers. The first linear layer compressed the dimension to a hidden size of 512 neurons, following which the output was normalized (using LayerNorm) and transformed with a SiLU nonlinearity. The second layer preserved the hidden dimension but also involved normalization and nonlinearity. The final layer condensed the features to the latent space size. 

Both the encoders described above were subsequently used with a decoder comprising three linear layers interspersed with a normalization layer (LayerNorm) and SiLU nonlinearities as in the case of the nonlinear encoder, except that the dimensions sequentially increased from 128 to 512 to the output size. 

\begin{figure}
\includegraphics[width=1.0\linewidth]{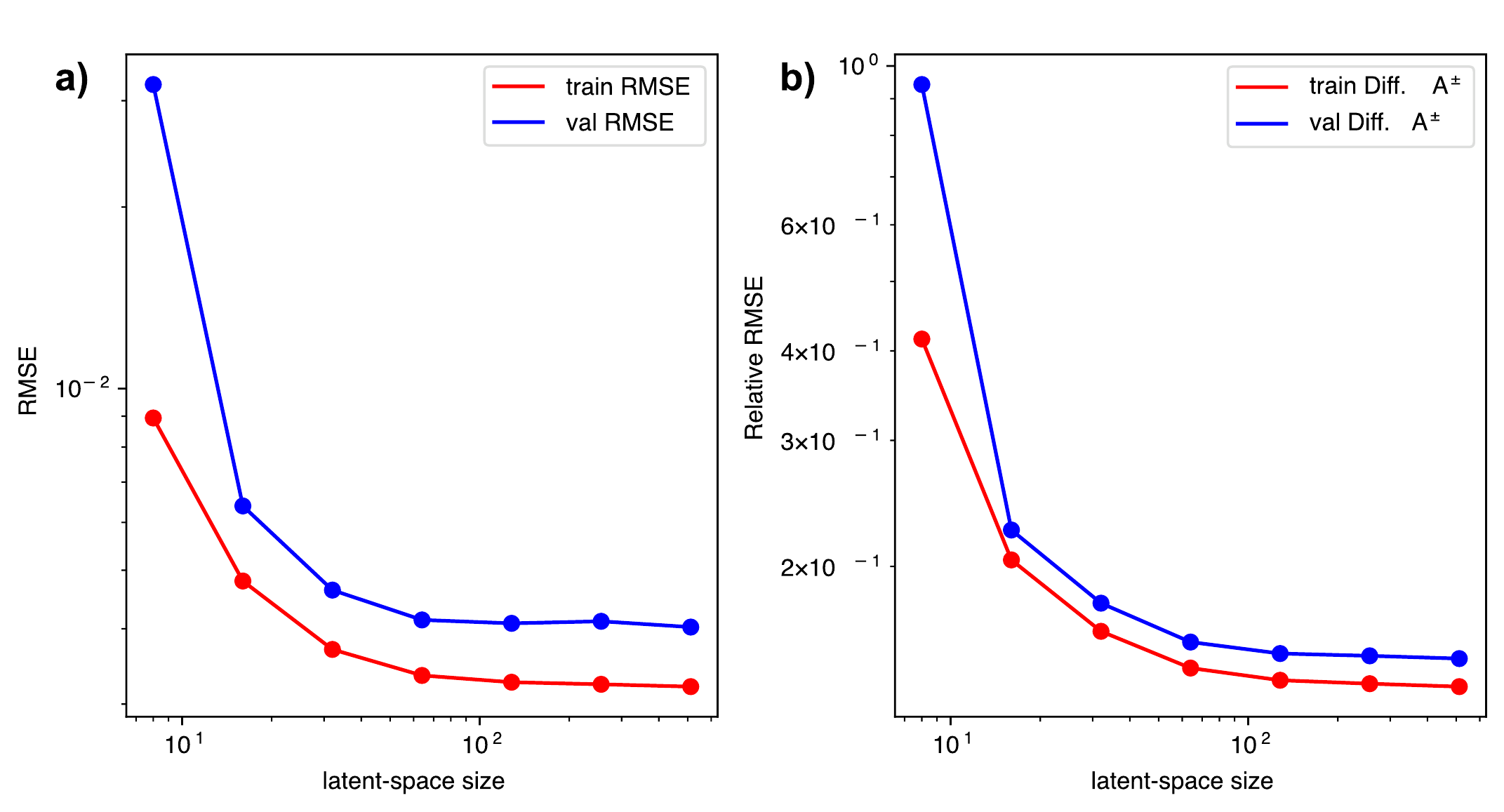}
\caption{
a) Convergence of overall RMSE in the prediction of the $\rep|\frho_i^7>$ features using encoder/decoder architectures with changing hidden layer dimension (latent space sizes) considering both structures in the degenerate pairs (A$^+$, A$^-$). b) Highlights, in particular, the convergence of relative RMSE in the difference of the features of the degenerate pairs.}
\label{fig:latent-space-error}
\end{figure}

The validation error in reconstructing $\rep|\frho_i^7>$ features converges quickly with respect to the latent-space size, suggesting that a relatively small number of descriptors is necessary to fully characterize the environment structure in a rotationally-invariant fashion (Fig.~\ref{fig:latent-space-error}). 

\subsection{Details of energy models}
We consider as targets, the difference between the energy of the structure and the one obtained for the minimum energy structure (a centered heptagon). Just as in the case of feature models, we considered linear and nonlinear models.
The linear models consist of two linear layers sequentially reducing the input features to the a latent-space size of 128 features and finally to a single scalar output per structure. We use this structure even though obviously a single weight vector would suffice to mimic the behavior of the contracted features model during training. 
Nonlinear models on the other hand first compress the features to a hidden size of 128 following which a SiLU nonlinearity is applied and the procedure is repeated for another intermediate layer before producing a single energy output. 

\subsection{Training Errors}
We report the errors incurred while training the various models described on a dataset of 3500 pairs of the \ce{B_8} clusters in Table ~\ref{tab:results-train}. For simplicity, we denote by the subscript L if the models employed were linear, or NL if the model was nonlinear.  A$_\text{L}^{\rho}$,  A$_\text{NL}^{\rho}$ with superscript $\rho$ indicates that the model built on three-center features was targeting the high $\nu$ atom-centered feature. 
The encoded features from A$_\text{L}^{\rho}$ and A$_\text{NL}^{\rho}$ models were subsequently used as inputs with the nonlinear energy models, yielding A$_\text{L}$, A$_\text{NL}$ models respectively.
If instead the energies were targeted directly by using the $N=2, \nu=1$ features, we obtain B$_\text{L}$ and B$_\text{NL}$ respectively. 
We also consider both linear and nonlinear energy models when using $\rep|\frho_i^7>$ as the inputs, C$_\text{L}$ and C$_\text{NL}$ respectively. Additionally, we construct D$_\text{L}$ and D$_\text{NL}$ that are linear and nonlinear energy models using $\rep|\frho_i^3>$ as the input. 
\begin{table}[btp]
    \centering
    \begin{tabular}{l r@{}c@{}l c c }
    \hline\hline
       \multicolumn{4}{c}{Model}    & \multicolumn{2}{c}{Relative Train Error}   \\
  Name &  \multicolumn{3}{c}{Architecture} & Mean $A^\pm$ & Diff. $A^\pm$ \\
    \hline   
A$_\text{L}^{\rho}$  &  $\rho_{ii_1i_2}^{\otimes 1}$$\substack{\longrightarrow\\\text{L}}$ & $\tilde{\feat}_p$  & $\substack{\longrightarrow\\\text{NL}}$ $\rho_i^{\otimes 7}$        &   0.009   &   1.00
   \\
A$_\text{L}$  &  & & $\!\!\!\!\substack{\hookrightarrow\\\text{NL}}$ $E$  & 0.028 & 1.00  \\
A$_\text{NL}^{\rho}$  &  $\rho_{ii_1i_2}^{\otimes 1}$$\substack{\longrightarrow\\\text{NL}}$ & $\tilde{\feat}_p$  & $\substack{\longrightarrow\\\text{NL}}$ $\rho_i^{\otimes 7}$        &   0.0043      &    0.125
   \\
A$_\text{NL}$  &  & & $\!\!\!\!\substack{\hookrightarrow\\\text{NL}}$ $E$  & 0.001 & 0.0481 \\
B$_\text{L}$  &  \multicolumn{3}{c}{$\rho_{ii_1i_2}^{\otimes 1}$$\substack{\longrightarrow\\\text{L}}$  \!$E$\!  }  & 0.417    & 1.00 \\
B$_\text{NL}$  &  \multicolumn{3}{c}{$\rho_{ii_1i_2}^{\otimes 1}$$\substack{\longrightarrow\\\text{NL}}$  \!$E$\!  }  &  0.005  & 0.107 \\
C$_\text{L}$ &   \multicolumn{3}{c}{$\rho_{i}^{\otimes 7}$$\substack{\longrightarrow\\\text{L}}$  \!$E$\!  }  &  0.074   & 0.550 \\
C$_\text{NL}$ &   \multicolumn{3}{c}{$\rho_{i}^{\otimes 7}$$\substack{\longrightarrow\\\text{NL}}$  \!$E$\!  }  &  0.045 & 0.211 \\
D$_\text{L}$ &   \multicolumn{3}{c}{$\rho_{i}^{\otimes 3}$$\substack{\longrightarrow\\\text{L}}$  \!$E$\!  }  &  0.410 & 1.00 \\
D$_\text{NL}$ &   \multicolumn{3}{c}{$\rho_{i}^{\otimes 3}$$\substack{\longrightarrow\\\text{NL}}$  \!$E$\!  }  &  0.036 & 1.00 \\
    \hline\hline
\end{tabular}    
\caption{Accuracy of different models for learning the high-$\nu$ features, and the energy of the train set of 3500 bispectrum-degenerate pairs of B$_8$ structures. Similar to the main text, we report separately the error on the mean of the features and energies of degenerate pairs $(A^+,A^-)$, and on their difference, normalized by the standard deviation of the train set. Absolute energy errors can be obtained by multiplying the values by 3.84 eV (Mean $A^\pm$) and 0.250 eV (Diff. $A^\pm$).}
\label{tab:results-train}
\end{table}
\end{document}